# Structure and Thermal Conductivity of Epoxy–Graphene Oxide Composites from 4.5 to 300 K

A. Jeżowski[1], A.I. Krivchikov[1,2], O.A. Korolyuk[2], V.V. Sagan[2], Y.V. Horbatenko[2], S.V. Cherednychenko[2], D. Szewczyk[1]*

[1] Institute of Low Temperature and Structure Research, Polish Academy of Sciences, Wrocław, Poland

[2] B. Verkin Institute for Low Temperature Physics and Engineering, National Academy of Sciences of Ukraine, Kharkiv, Ukraine

* Corresponding author: d.szewczyk@intibs.pl

## Abstract

The microscopic mechanisms of heat transport in structurally disordered polymer–carbon composites remain incompletely understood, particularly over the broad temperature range extending from cryogenic to ambient conditions. Here, we investigate the relationship between structural organization and thermal transport in epoxy–graphene oxide (GO) composites with nominal GO contents of 0.25–10 wt.% relative to the resin (0.21–6.1 wt.% of the total composite mass). X-ray diffraction reveals predominantly amorphous scattering without a detectable GO(001) reflection, suggesting substantial suppression of ordered GO stacking within the cured epoxy matrix. Raman spectroscopy confirms the increasing contribution of the carbon phase while revealing no pronounced changes in the characteristic vibrational bands of the epoxy network. Thermal-conductivity measurements (4.5–300 K) demonstrate that GO incorporation modifies the temperature dependence of heat transport rather than producing a simple conductivity enhancement. The experimental data are described by a phenomenological multichannel model comprising a temperature-independent plateau term and two or three effective Arrhenius-type contributions characterized by distinct temperature scales ranging from a few to several hundred kelvin. The low-temperature conductivity plateau observed at lower GO concentrations is no longer resolved in GO-6 and GO-10, indicating a qualitative change in the temperature dependence of heat transport. These results establish a correlation between structural organization and the redistribution of effective thermal-transport contributions in epoxy–GO composites, providing an experimental basis for further microscopic interpretation within the framework of unified vibrational thermal-transport theory.

## 1. Introduction

Polymer composites with tailored thermal-transport properties are of considerable interest both for fundamental studies of heat conduction in structurally disordered materials and for applications in electronics, aerospace technologies, cryogenic engineering, and energy systems. Among thermosetting polymers, epoxy resins are widely used owing to their mechanical performance, chemical resistance, dimensional stability, and processing versatility. However, their low thermal conductivity, typically 0.1–0.5 W $m^{-1}$ $K^{-1}$ at room temperature, limits heat dissipation and motivates the development of composites with controllable thermal-transport properties [1, 2].

A widely explored strategy is the incorporation of carbon-based fillers, particularly graphene and its derivatives [2–4]. Graphene combines exceptionally high intrinsic thermal conductivity with a large specific surface area and high aspect ratio, which can promote the formation of thermally conductive pathways within the polymer matrix [3, 4]. Graphene oxide (GO) offers additional advantages owing to its oxygen-containing functional groups, which improve wettability, dispersibility, and interfacial interactions with epoxy resins [5, 6]. At the same time, oxidation disrupts the $sp^2$-bonded carbon network and introduces structural defects, thereby reducing the intrinsic thermal conductivity of the carbon phase. Consequently, GO cannot be regarded simply as a highly conductive inclusion embedded in a poorly conducting matrix.

The effective thermal conductivity of epoxy–GO composites is governed by the interplay among filler concentration, morphology, orientation, dispersion, agglomeration, network connectivity, degree of reduction, and thermal coupling at the filler–matrix interfaces [1–4, 7, 8]. Interconnected carbon networks can facilitate heat transfer, whereas structural defects, agglomeration, and interfacial thermal resistance may suppress it. In particular, graphene–polymer interfaces can contribute substantially to the overall thermal resistance [7, 8]. The incorporation of GO may therefore modify not only the magnitude of thermal conductivity but also the mechanisms governing its temperature dependence.

From a microscopic perspective, epoxy–GO composites are heterogeneous disordered systems in which an amorphous polymer matrix coexists with locally ordered, defect-rich carbon domains and a high density of interfaces. In crystalline solids, heat conduction is commonly described within the phonon-gas framework, where vibrational lifetimes are limited by anharmonic interactions, structural defects, and boundary scattering [9]. In amorphous solids, strongly scattered, diffusive, and non-propagating vibrational excitations can make substantial contributions to thermal transport [10–13]. These distinct transport regimes are reflected in the characteristic temperature dependence of thermal conductivity. Whereas ordered crystals typically exhibit a pronounced phonon-conductivity maximum, structurally disordered solids often display a weakly temperature-dependent conductivity plateau at cryogenic temperatures, followed by a renewed increase upon heating [10, 12, 13]. The evolution of these features provides insight into the influence of structural disorder and vibrational scattering on heat transport. In heterogeneous polymer–carbon composites, the incorporation of GO may further modify this behavior through changes in local structural organization, interfacial coupling, and the relative contributions of different heat-transport processes.

The unified theory of thermal transport developed by Simoncelli, Marzari, and Mauri within the Wigner formulation expresses thermal conductivity as the sum of a population (particle-like) contribution and a coherence (wave-like) contribution arising from correlations between vibrational modes [14–16, 27, 28]. This framework provides a unified description of the crossover between crystal-like and glass-like thermal transport and suggests that, in structurally complex materials such as epoxy–GO composites, heat conduction need not be governed by a single class of vibrational excitations. A quantitative separation of these contributions requires microscopic information on the vibrational spectrum and mode-resolved transport properties. Nevertheless, experimentally observed temperature dependences and characteristic temperature scales may provide useful constraints on the underlying transport processes and their sensitivity to structural modifications.

The importance of investigating the complete temperature dependence of thermal conductivity is further highlighted by Nataj et al. [23], who showed that graphene-based composites exhibiting enhanced thermal conductivity near room temperature may behave as comparatively good thermal insulators under cryogenic conditions. Despite extensive studies of graphene- and GO-based polymer composites [1–8], systematic investigations extending from cryogenic temperatures to 300 K and combining structural characterization with thermal-conductivity measurements on the same samples remain relatively scarce.

In this work, we investigate a series of epoxy–GO composites with nominal GO contents of 0.25–10 wt.% relative to the resin (corresponding to 0.21–6.1 wt.% of the total composite mass) using X-ray diffraction, Raman spectroscopy, and thermal-conductivity measurements over the temperature range 4.5–300 K. We examine how GO incorporation affects the structural organization of the epoxy matrix and whether it modifies the temperature dependence of thermal conductivity rather than merely its magnitude. Particular attention is given to the evolution of the low-temperature conductivity plateau and to the characteristic temperature scales identified through a phenomenological multichannel representation.

## 2. Experimental

### 2.1 Sample preparation

Graphene oxide (GO) was synthesized from high-purity graphite using a modified Hummers method developed at the B. Verkin Institute for Low Temperature Physics and Engineering of the National Academy of Sciences of Ukraine [24]. The oxidation process introduces oxygen-containing functional groups, disrupts the $sp^2$-bonded carbon network, and increases the interlayer spacing [25]. Thermally reduced graphene oxide (trGO-300), obtained by reduction at 300 °C, was used as a structural and spectroscopic reference for the carbon phase [24].

Epoxy–GO composites were prepared by dispersing GO powder in the epoxy resin (component A). A predetermined amount of GO was added to component A, followed by mechanical stirring until a macroscopically homogeneous mixture was obtained. The mixture was subsequently sonicated for 30 min at a frequency of 22 kHz. The curing agent (component B) was then added at the mass ratio specified in Table 1. After further mixing, the resulting mixture was cast into molds and allowed to cure. Two neat-epoxy reference samples were prepared using different curing-agent-to-resin mass ratios and are denoted as pure20 and pure70.

The final GO content was calculated relative to the total mass of the composite (resin + GO + curing agent). Throughout this work, the composite samples are designated GO-0.25, GO-0.5, GO-1, GO-3, GO-6, and GO-10 according to their nominal GO content relative to component A. The corresponding GO mass fractions in the final composites are listed in Table 1.

**Table 1.** Composition of the epoxy–GO formulations.

| Sample | Nominal GO rel. to A, wt.% | $m_A$, g | $m_{GO}$, g | $m_B$, g | $m_B/m_A$, % | Total, g | Final GO, wt.% |
|---|---|---|---|---|---|---|---|
| GO-0.25 | 0.25 | 28.000 | 0.070 | 5.600 | 20.00 | 33.670 | 0.208 |
| GO-0.5 | 0.50 | 21.000 | 0.105 | 4.200 | 20.00 | 25.305 | 0.415 |
| GO-1 | 1.00 | 14.000 | 0.140 | 2.800 | 20.00 | 16.940 | 0.826 |
| GO-3 | 3.00 | 7.000 | 0.210 | 2.100 | 30.00 | 9.310 | 2.256 |
| GO-6 | 6.00 | 7.000 | 0.420 | 2.100 | 30.00 | 9.520 | 4.412 |
| GO-10 | 10.00 | 30.000 | 3.000 | 16.300 | 54.33 | 49.300 | 6.085 |

## 2.2 Structural characterization

Powder X-ray diffraction (XRD) patterns were recorded using Cu Kα radiation ($\lambda = 1.5406$ Å) over the range $2\theta = 3$–$100°$, with a step size of approximately 0.0066°. The diffraction patterns were smoothed using the Savitzky–Golay method, and the main amorphous halo was fitted with Gaussian profiles over the range $2\theta = 15$–$28°$. The corresponding interplanar spacings were calculated using Bragg's law. The increase in scattering intensity below $2\theta \approx 8$–$10°$, observed in all samples, including the neat epoxy references, was treated as parasitic low-angle scattering and excluded from the analysis.

Raman spectra were recorded using a Renishaw inVia Raman microscope equipped with a Leica DM 2500 confocal system and a 514 nm excitation laser over the spectral range 100–4000 $cm^{-1}$. The D and G bands of trGO-300 were fitted using two Lorentzian profiles and a linear background over the range 1000–1800 $cm^{-1}$. In the epoxy–GO composites, the *G* band overlaps with the aromatic ring stretching band of the epoxy matrix at approximately 1608 cm$^{-1}$. Therefore, the relative prominence of the *D* band was evaluated semi-quantitatively from its height at approximately 1355 cm$^{-1}$ above a linear baseline defined over 1260–1480 cm$^{-1}$, normalized to the height of the band at 1608 cm$^{-1}$.

### 2.3 Thermal conductivity

Thermal conductivity was measured over the temperature range 4.5–300 K using the steady-state uniaxial heat-flow method in a custom-built apparatus [26]. Measurements were performed under high-vacuum conditions, with the specimen surrounded by multiple radiation shields. The shield closest to the sample was actively temperature-controlled to minimize radiative heat exchange.

A heat flux was generated by a miniature resistive heater attached to one end of the specimen, while the temperature difference was measured using calibrated constantan–manganin differential thermocouples separated by approximately 10 mm. The total measurement uncertainty did not exceed 15%, including an estimated contribution of approximately 5% from the uncertainty in thermocouple spacing. The measurement reproducibility was better than ±2%.

## 3. Results

### 3.1 Structure: XRD

All epoxy-based samples exhibit a broad amorphous halo near $2\theta \approx 21°$ in the range shown in Fig. 1, without sharp diffraction reflections. An additional broad scattering feature near $2\theta \approx 43°$ is observed in the full-range XRD data. The low-angle intensity increase below approximately 8–10° is excluded from the structural analysis, as described in Section 2.2. Notably, the GO(001) reflection at $2\theta \approx$ 10–13° ($d \approx$ 0.7–0.9 nm), characteristic of ordered GO stacking, is not detected at any GO loading. This observation suggests substantial suppression of ordered GO stacking within the cross-linked epoxy matrix, although the absence of this reflection alone does not establish complete exfoliation of the GO sheets.

The position of the main amorphous halo varies with GO content (Table 2, Fig. 2). For nominal GO contents up to 6 wt.% relative to the epoxy resin, the main halo is located at lower diffraction angles than in neat epoxy, corresponding to characteristic real-space distances of approximately 4.6–4.9 Å, compared with 4.35 Å for the neat matrix. The halo width remains relatively unchanged (FWHM = 7.6–8.7°). These changes are consistent with a modest increase in the characteristic intermolecular spacing within the epoxy matrix, potentially associated with the incorporation of dispersed GO sheets. In GO-10, the main halo is located at a higher diffraction angle than in the other composites ($d$ = 4.12 Å) and exhibits appreciable broadening (FWHM = 9.2°). The two neat-epoxy references, prepared with different curing-agent-to-resin ratios, exhibit similar halo positions (d = 4.22–4.35 Å). The parameters of pure20 are presented in Table 2 as a representative reference, while pure70 is retained in Fig. 1 for comparison.

**Table 2.** Parameters of the main amorphous XRD halo in neat epoxy and epoxy–GO composites.

| Sample | $2\theta c$, ° | FWHM, ° | *d*, Å |
|---|---|---|---|
| Neat epoxy (pure20) | 20.4 | 7.6 | 4.35 |
| GO-0.25 | 18.6 | 8.0 | 4.77 |
| GO-0.5 | 19.4 | 7.7 | 4.57 |
| GO-1 | 18.3 | 8.0 | 4.85 |
| GO-3 | 18.2 | 7.9 | 4.88 |
| GO-6 | 18.4 | 8.7 | 4.82 |
| GO-10 | 21.6 | 9.2 | 4.12 |

The trGO-300 reference exhibits no detectable GO(001) reflection and shows a broad turbostratic scattering feature at $2\theta \approx 17$–$24°$ (Fig. 1). The full-range XRD data also reveal a broad in-plane (100)/(101) reflection at $2\theta \approx 43.3°$ ($d \approx 2.09$ Å), consistent with retained in-plane $sp^2$ ordering and disordered interlayer stacking. The broadening of the GO-10 amorphous halo may suggest changes in the structural organization of the carbon filler, including possible partial restacking. However, this interpretation remains tentative because GO-10 also contains a substantially higher curing-agent fraction (Table 1), and changes in the epoxy network structure may contribute to the observed diffraction response.

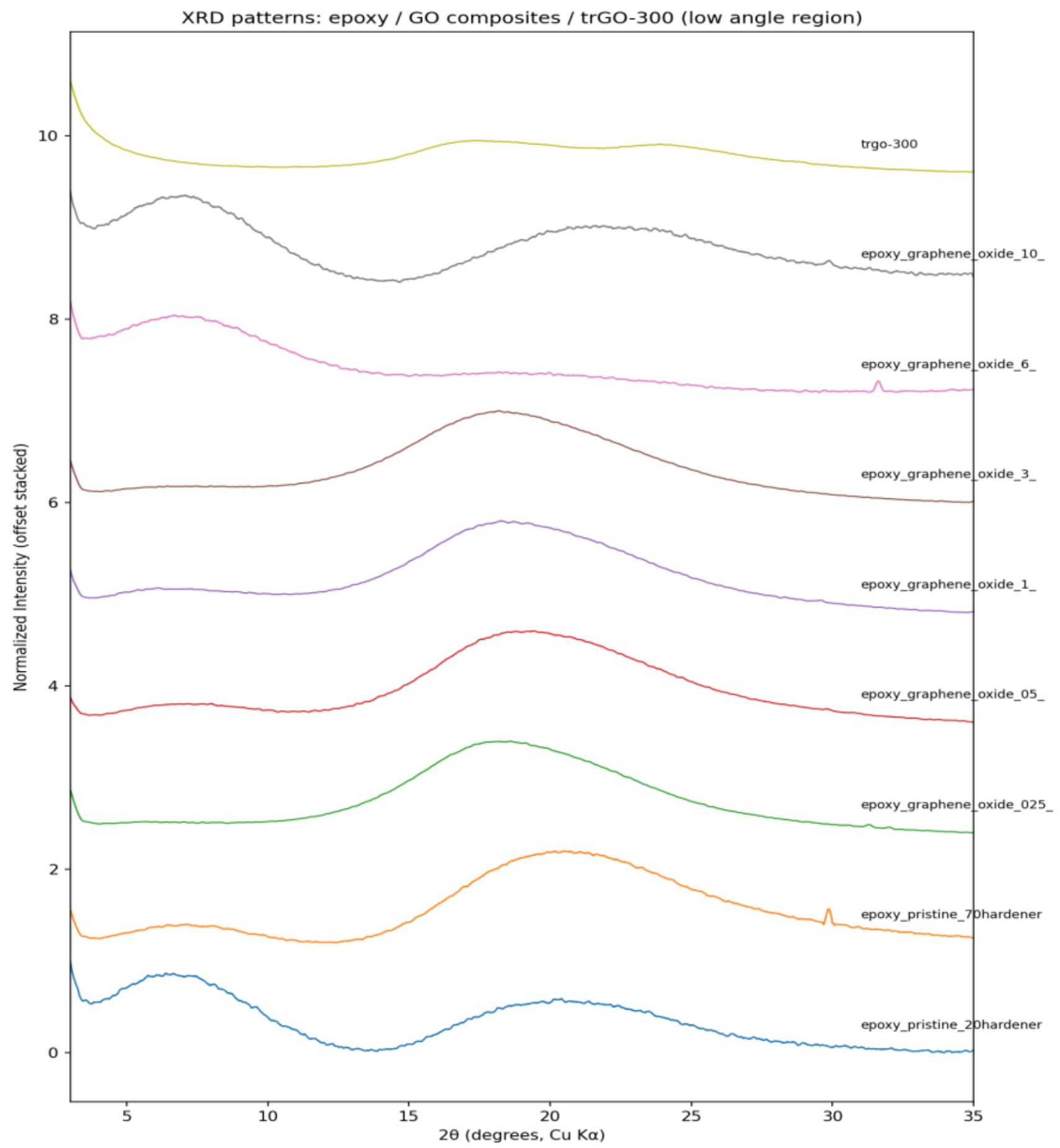

**Fig. 1.** Normalized and vertically offset XRD patterns of the neat epoxy references (pure20 and pure70), epoxy–GO composites (GO-0.25 to GO-10), and thermally reduced graphene oxide (trGO-300), recorded using Cu Kα radiation over the range 2$\theta$ = 3–35°.

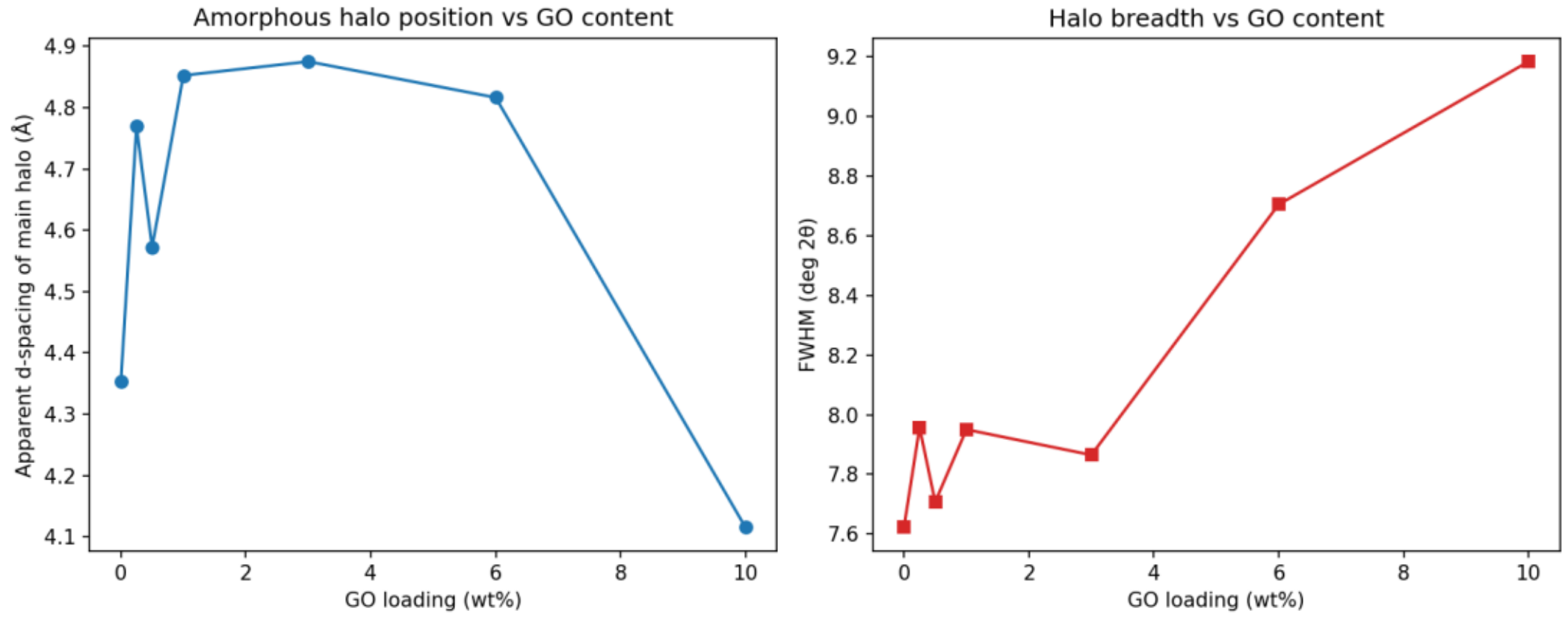

**Fig. 2.** Characteristic real-space distance, *d* (left), and full width at half maximum, FWHM (right), of the main amorphous XRD halo as functions of the nominal GO content relative to the epoxy resin.

### 3.2 Structure: Raman spectroscopy

Neat epoxy exhibits characteristic molecular Raman bands of the bisphenol-A-based network, including $CH_2/CH_3$ stretching vibrations near 2870–2940 $cm^{-1}$ and aromatic C–H stretching vibrations at approximately 3070 $cm^{-1}$ (Fig. 3). The trGO-300 reference exhibits the characteristic *D* and *G* bands of reduced graphene oxide at 1364 and 1596 $cm^{-1}$, respectively, with corresponding full widths at half maximum (FWHM) of 174 and 87 $cm^{-1}$. The intensity ratio $I_D/I_G$ is approximately 0.87 when calculated from peak heights and 1.74 when calculated from integrated peak areas, consistent with a defect-rich $sp^2$ carbon network.

In the epoxy–GO composites, a *D*-band shoulder at approximately 1350–1365 $cm^{-1}$, absent in neat epoxy, becomes clearly distinguishable from GO-1 onwards. The *D*-band prominence relative to the epoxy band at 1608 $cm^{-1}$ increases from 0.26 for GO-1 to 0.42 for GO-3 and 0.80 for GO-6 (Table 3). For GO-0.25 and GO-0.5, this parameter is unreliable because the sharp epoxy bands near 1300 and 1460 $cm^{-1}$ interfere with baseline determination. For GO-10, the overall Raman signal is substantially weaker, by approximately one to two orders of magnitude, potentially due to increased optical absorption by the carbon phase. Consequently, the lower *D*/1608 ratio of 0.65 should be interpreted with caution because of the reduced signal-to-noise ratio. The characteristic Raman bands of the epoxy matrix, including those in the C–H stretching region, show no pronounced changes up to GO-6, suggesting that no major modifications of the epoxy vibrational fingerprint are detected within the sensitivity of the measurements.

**Table 3.** Relative prominence of the *D* band normalized to the epoxy Raman band at 1608 $cm^{-1}$. Negative values indicate that no *D*-band contribution could be reliably resolved.

| Sample | *D*/1608 |
|---|---|
| pure20 / pure70 | −0.17 / −0.20 |
| GO-0.25 | 0.10 (unreliable) |
| GO-0.5 | −0.01 (unreliable) |
| GO-1 | 0.26 |
| GO-3 | 0.42 |
| GO-6 | 0.80 |
| GO-10 | 0.65 (low S/N) |

Taken together, the XRD and Raman results are consistent with the incorporation of GO into the epoxy matrix without pronounced changes in its Raman vibrational fingerprint. The diffraction data also suggest suppression of ordered GO stacking, although partial restacking at the highest GO loading cannot be excluded.

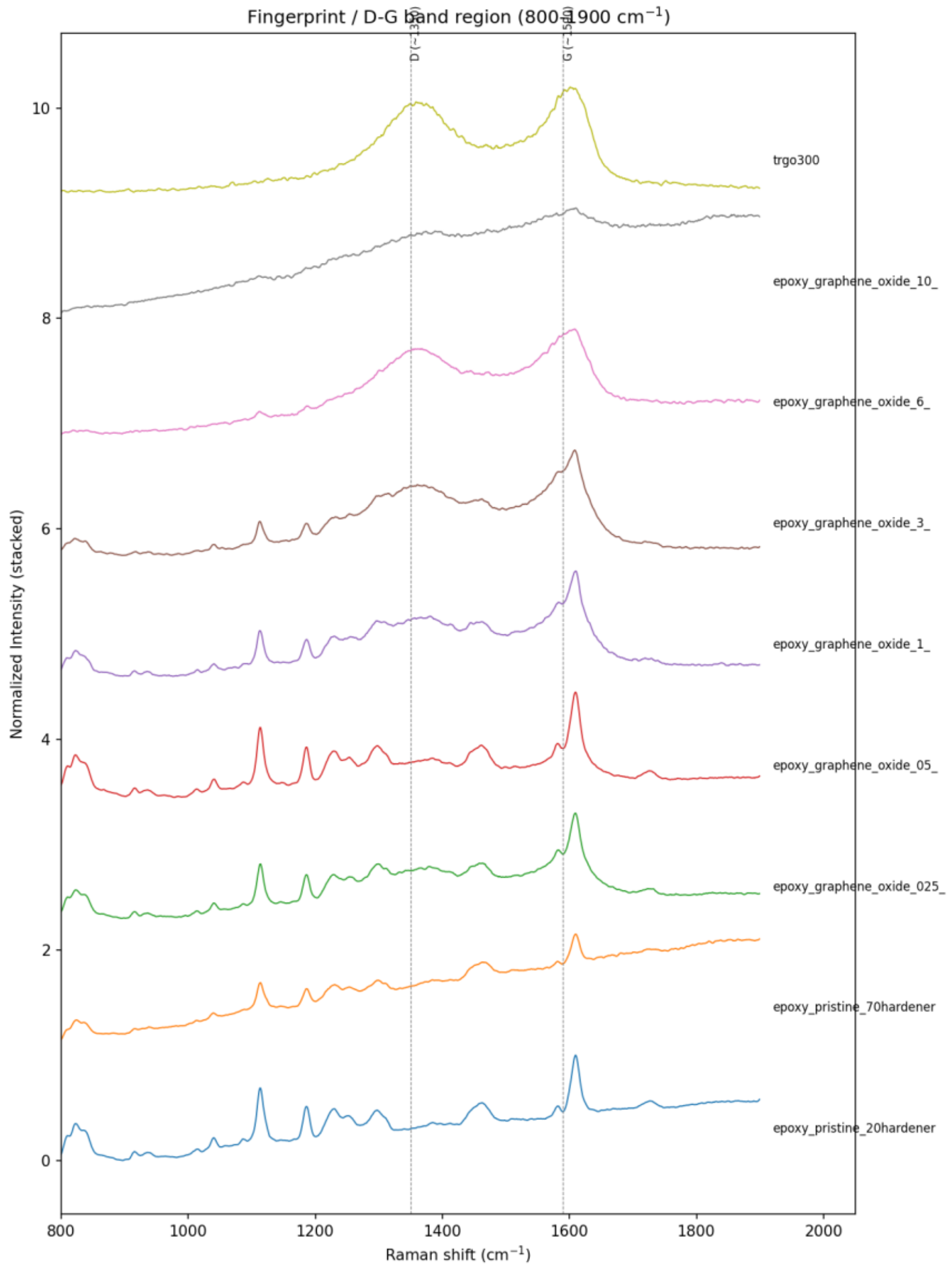

**Fig. 3.** Normalized and vertically offset Raman spectra of the neat epoxy references (pure20 and pure70), epoxy–GO composites (GO-0.25 to GO-10), and trGO-300 in the spectral range 800–1900 $cm^{-1}$. Dashed lines indicate the reference positions of the *D* (≈1350 $cm^{-1}$) and *G* (≈1590 $cm^{-1}$) bands.

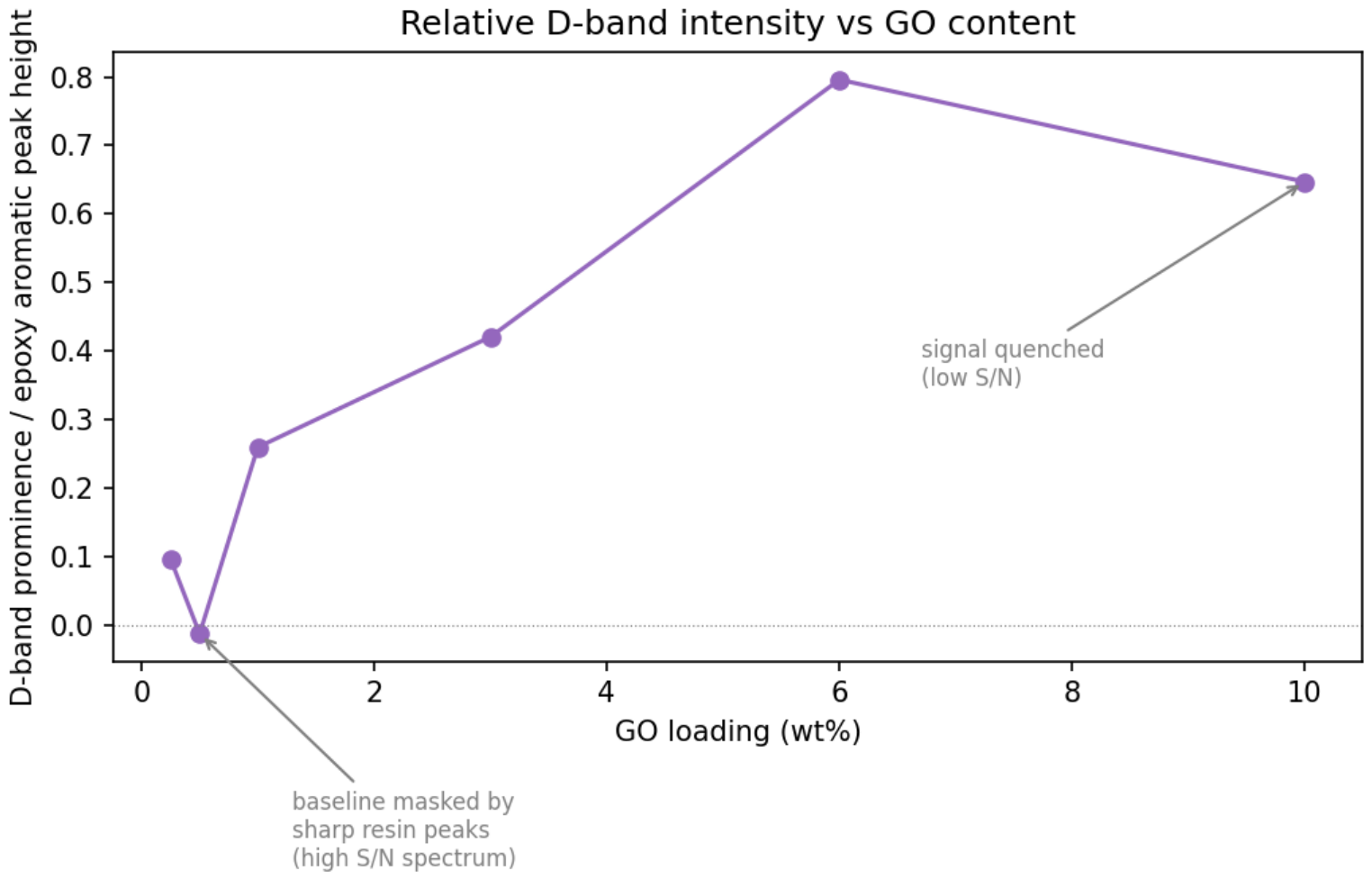

**Fig. 4.** Relative *D*-band prominence normalized to the epoxy Raman band at 1608 $cm^{-1}$ as a function of nominal GO content relative to the epoxy resin. The slightly negative value obtained for GO-0.5 arises from baseline-estimation uncertainty and does not represent a physically negative *D*-band intensity.

## 3.3 Thermal conductivity

### 3.3.1 Crystal-like and glass-like thermal transport

Ordered crystals usually exhibit a bell-shaped $\kappa(T)$ dependence with a pronounced phonon-conductivity maximum, commonly described within the Peierls–Boltzmann framework [33]. Glasses, by contrast, typically show $\kappa \propto T^2$ below approximately 1 K, a plateau near 5–15 K, and a renewed increase at higher temperatures [31, 32]. These features have been discussed in terms of resonant scattering by two-level systems and soft quasi-localized vibrations [29, 30, 34–36], as well as diffuson-mediated transport [12, 13]. The Wigner formulation provides a unified description in terms of population (particle-like) and coherence (wave-like) contributions [14–16]:

$$\kappa(T) = \kappa_P(T) + \kappa_C(T). \quad (3)$$

In disordered solids, the coherence contribution can become increasingly important as vibrational modes broaden and overlap, providing a connection to the diffuson picture of Allen and Feldman [12, 13, 17, 18]. Following our earlier work [19–22, 37–39], we write

$$\kappa(T) = [(\beta T^n)^{-1} + (A/T + \kappa_{pl})^{-1}]^{-1} + \kappa_0\, exp(-E/T), \quad (4)$$

where the first term phenomenologically combines the low-temperature power-law contribution with a higher-temperature contribution containing an $A/T$ dependence and a plateau term $\kappa_{pl}$, using an inverse-addition form analogous to Matthiessen's rule [40]. The second term represents an effective Arrhenius-type contribution, where $\kappa_0$ denotes its formal high-temperature limiting amplitude and $E$ is a characteristic temperature scale expressed in kelvin. Although this phenomenological form can be considered in the context of unified vibrational thermal-transport theory, it does not constitute a direct quantitative determination of the coherence contribution within the Wigner formulation.

For the epoxy–GO composites, no crystal-like maximum is observed. Moreover, because no thermal-conductivity data below 4.5 K are available, the low-temperature power-law contribution $\beta T^n$ cannot be reliably determined. The experimentally observed low-temperature plateau is therefore represented phenomenologically by a temperature-independent term $\kappa_{pl}$, without assigning it to a specific microscopic transport mechanism. As shown below, a single Arrhenius term is insufficient to describe the measured temperature dependence. Following earlier phenomenological approaches to multichannel thermal transport in molecular crystals and their solutions [38] and carbon allotropes [41], we represent the measured thermal conductivity as a sum of a temperature-independent plateau term and up to three effective Arrhenius-type contributions. These terms are introduced as fitting components and should not be identified directly with individual microscopic transport mechanisms or with the population and coherence contributions of the Wigner formulation:

$$\kappa(T) = \kappa_{pl} + \Sigma_i \, \kappa_{0i} \, exp(-E_i/T), \text{ i} = 1,\ldots,3. \quad (5)$$

The parameters of each channel were initially estimated successively from approximately linear regions of the Arrhenius representation $ln(\kappa - \kappa_{pl})$ versus $1/T$, followed by refinement of Eq. (5) over the full temperature range.

### 3.3.2 Overview

Figure 5 compares $\kappa(T)$ of samples GO-0.25, GO-3, GO-6 and GO-10 with literature data for neat epoxies [23, 42–44] and GO-based materials [41, 45, 46]. The thermal conductivity of all composites remains within the range typical of epoxy resins and increases monotonically with temperature, while its temperature dependence differs markedly from that of the carbon reference materials. A low-temperature plateau characteristic of glass-like thermal transport is discernible for GO-0.25 and GO-3 but is not clearly resolved for GO-6 and GO-10 within the measured temperature range. Data for the intermediate compositions (GO-0.5 and GO-1) closely follow those of GO-0.25 and are omitted for clarity.

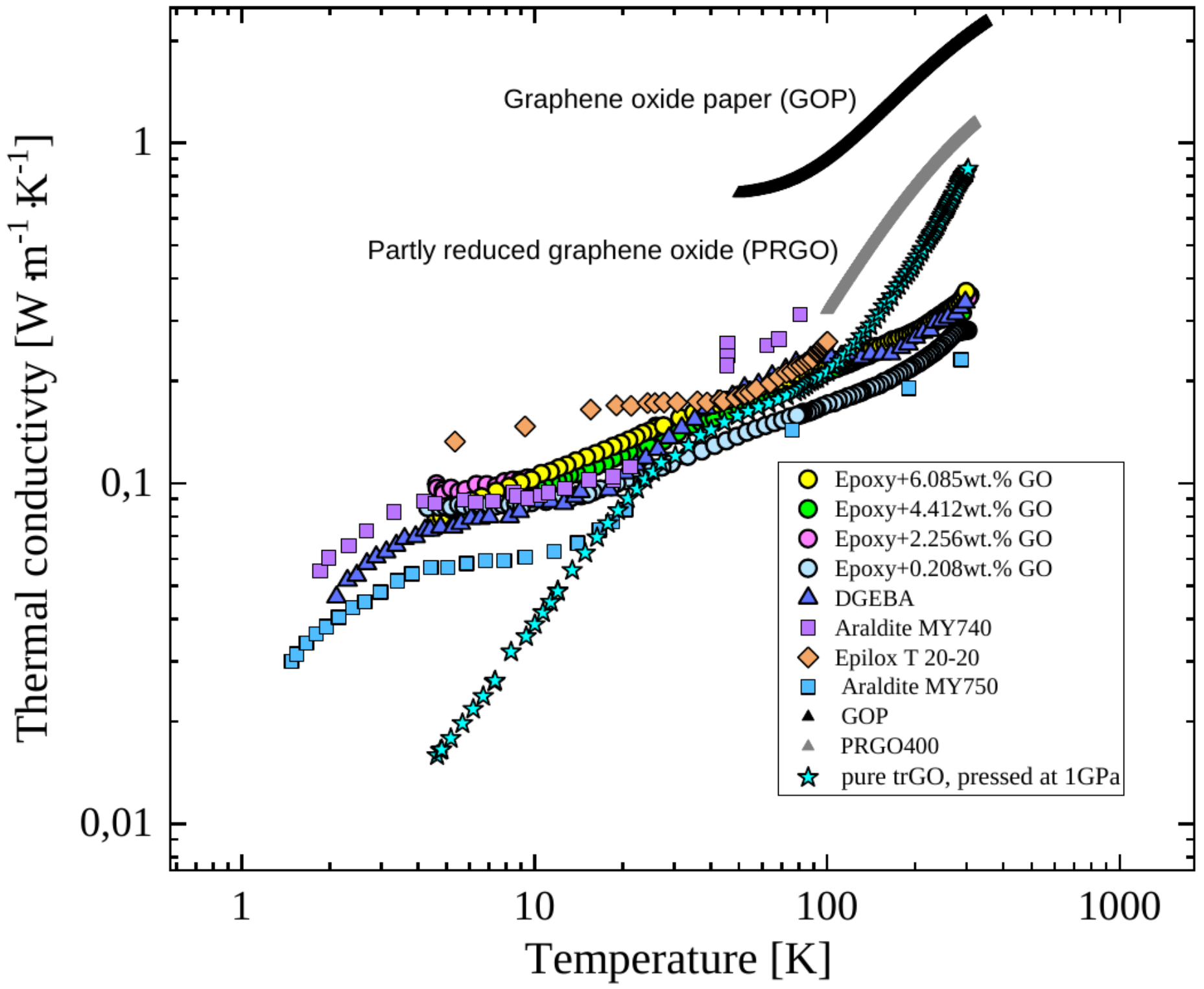

**Fig. 5.** Temperature dependence of the thermal conductivity $\kappa(T)$ of epoxy–GO composites GO-0.25, GO-3, GO-6 and GO-10. Literature data for neat epoxy resins (DGEBA [23], Araldite MY740 [42], Epilox T 20-20 [43] and Araldite MY750 [44]), graphene oxide paper (GOP [45]) and partially reduced graphene oxide (PRGO [46]) are included for comparison. Data for pure thermally reduced graphene oxide (trGO) compacted at 1 GPa [41] are also shown as a reference for the carbon phase. The measurement uncertainty is below 15%.

### 3.3.3 Low GO content (GO-0.25 and GO-3)

For both compositions, $\kappa(T)$ exhibits weakly temperature-dependent, plateau-like regions below approximately 10 K (Fig. 6), with a slightly more pronounced temperature dependence for GO-3. Over the entire measured temperature range, the thermal conductivity of GO-3 exceeds that of GO-0.25 by approximately 12% at $T \leq 7$ K and 21% at 300 K. A single activated channel describes the data only up to approximately 70 K for GO-0.25 and 30 K for GO-3. Including a second channel extends the description to 300 K for GO-0.25 and approximately 120 K for GO-3, while a third channel is required to describe the GO-3 data up to 300 K.

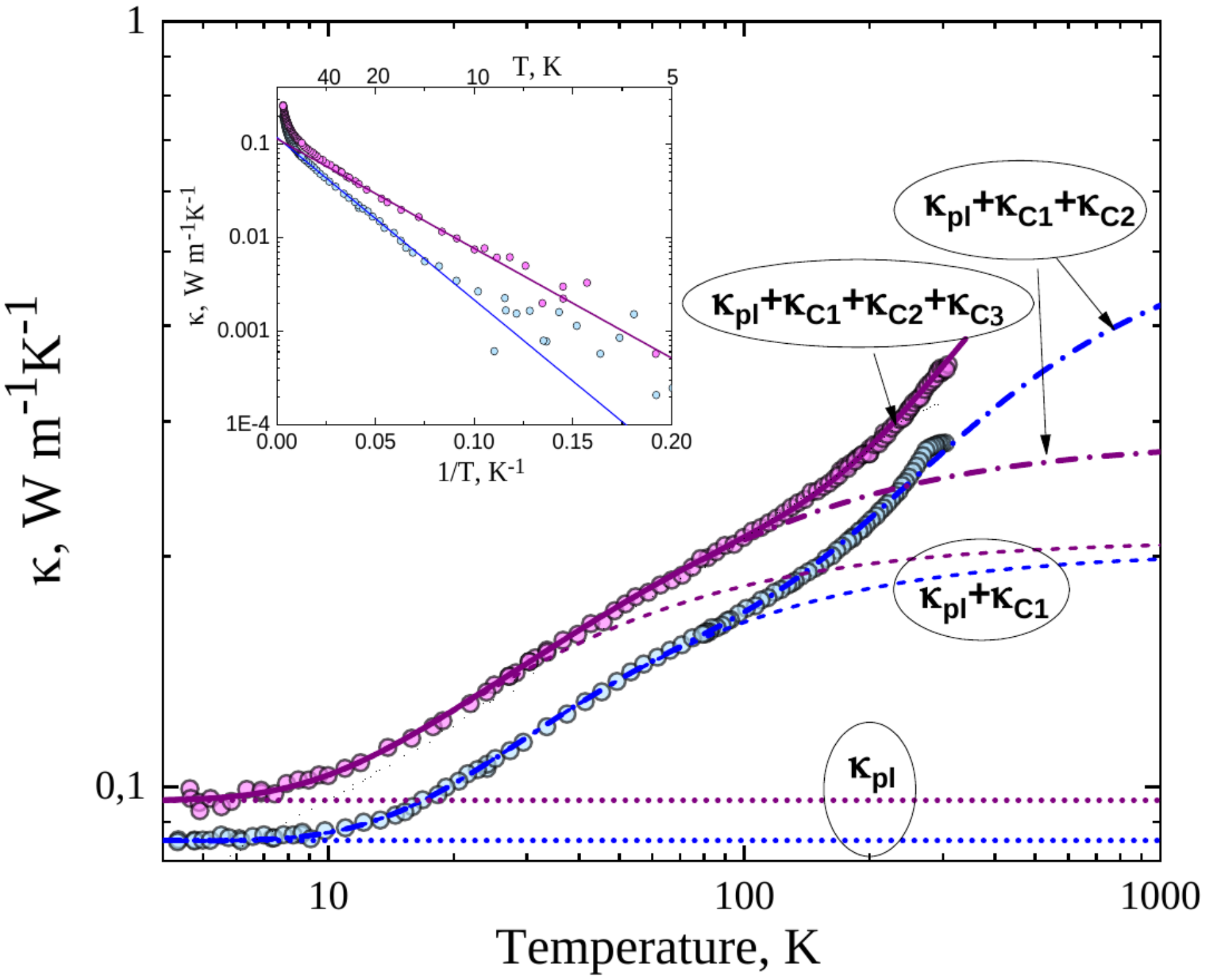

**Fig. 6.** Temperature dependence of the thermal conductivity $\kappa(T)$ of epoxy–GO composites GO-3 (pink) and GO-0.25 (light blue). Symbols represent experimental data. Lines show the fits: dotted lines, $\kappa_{\mathrm{pl}}$; dashed lines, $\kappa_{\mathrm{pl}} + \kappa_{C1}(T)$; dash-dotted lines, $\kappa_{\mathrm{pl}} + \kappa_{C1}(T) + \kappa_{C2}(T)$; solid line, $\kappa_{\mathrm{pl}} + \kappa_{C1}(T) + \kappa_{C2}(T) + \kappa_{C3}(T)$. The colours of the lines correspond to those of the symbols. Inset: Arrhenius representation $ln(\kappa - \kappa_{\mathrm{pl}})$ versus $1/T$ with linear fits used to estimate $\kappa_{01}$ and $E_1$. Fit parameters are given in Table 4. The measurement uncertainty is below 15%.

### 3.3.4 High GO content (GO-6 and GO-10)

No distinct low-temperature plateau is observed within the measured temperature range (Fig. 7). The thermal conductivity increases monotonically with temperature, and the two compositions differ by only approximately 7% at 7 K, becoming nearly identical near 200 K. With $\kappa_{\mathrm{pl}} = 0$, a single activated channel describes the data only below approximately 11 K, two channels extend the description to approximately 90 K for GO-6 and 120 K for GO-10, and three channels reproduce the measured temperature dependence up to 300 K. For GO-6 and GO-10, the absence of a clearly resolved low-temperature plateau within the measured temperature range is represented phenomenologically by setting $\kappa_{\mathrm{pl}} = 0$. This fitting choice does not establish the physical absence of a plateau-like contribution or exclude its possible emergence at temperatures below the experimental range.

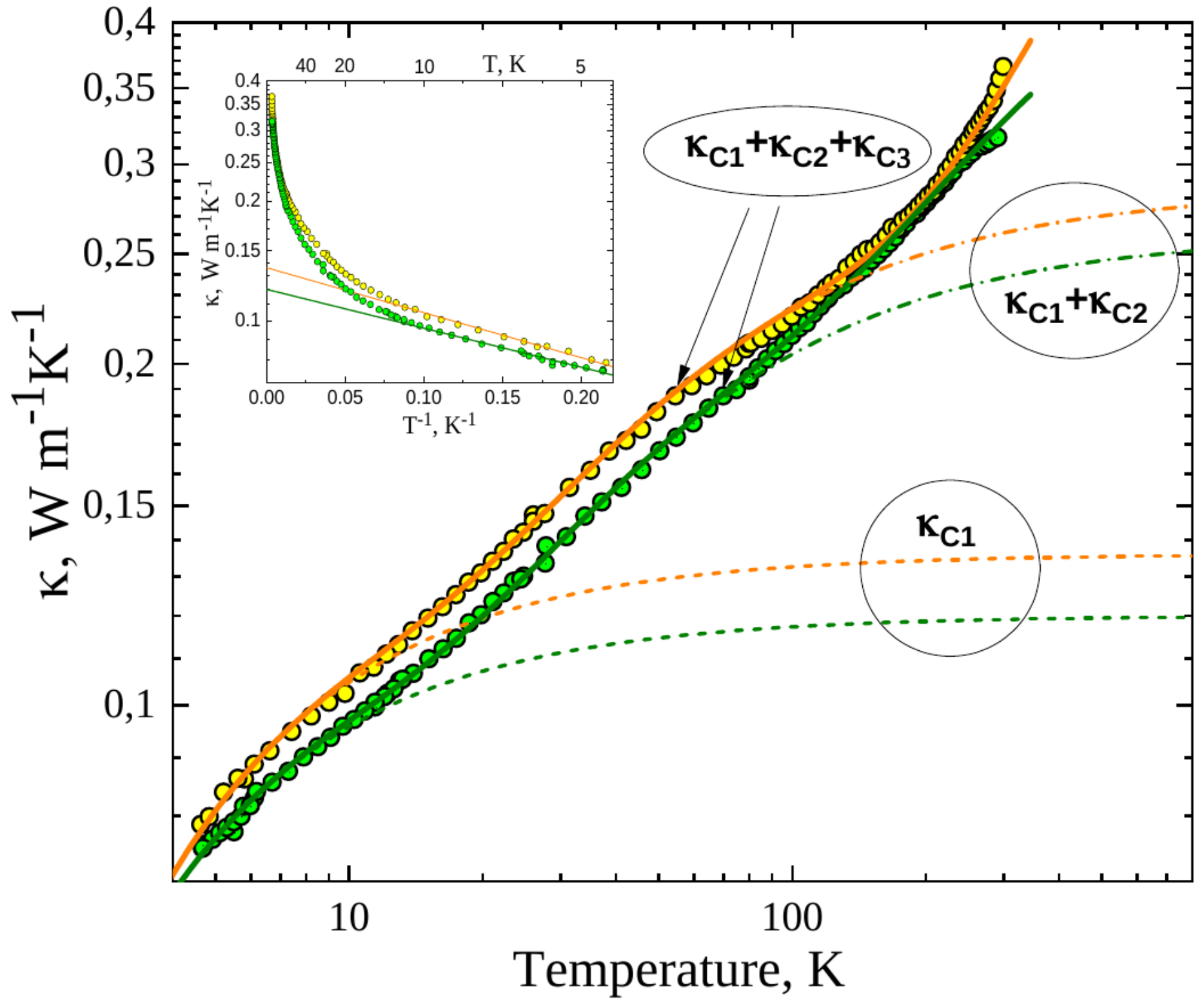

**Fig. 7.** Temperature dependence of the thermal conductivity $\kappa(T)$ of epoxy–GO composites GO-6 (green) and GO-10 (yellow). Symbols represent experimental data. Lines show the fits: dashed lines, $\kappa_{C1}(T)$; dash-dotted lines, $\kappa_{C1}(T) + \kappa_{C2}(T)$; solid lines, $\kappa_{C1}(T) + \kappa_{C2}(T) + \kappa_{C3}(T)$. The colours of the lines correspond to those of the symbols. Inset: Arrhenius representation of $\kappa$ versus $1/T$, with linear fits to the logarithm of $\kappa$ used to estimate $\kappa_{01}$ and $E_1$. Fit parameters are given in Table 4. The measurement uncertainty is below 15%.

**Table 4.** Parameters of Eq. (5). Thermal-conductivity coefficients $\kappa_{pl}$ and $\kappa_{0i}$ are expressed in W m$^{-1}$ K$^{-1}$, and characteristic energies $E_i$ in K.

| Sample | $\kappa_{pl}$ | $\kappa_{01}$ | $E_1$ | $\kappa_{02}$ | $E_2$ | $\kappa_{03}$ | $E_3$ |
|---|---|---|---|---|---|---|---|
| GO-0.25 | 0.085 | 0.118 | 40 | 0.346 | 420 | – | – |
| GO-3 | 0.096 | 0.114 | 27 | 0.074 | 100 | 0.75 | 600 |
| GO-6 | 0 | 0.12 | 2.25 | 0.14 | 48 | 0.301 | 365 |
| GO-10 | 0 | 0.136 | 2.6 | 0.149 | 50 | 0.86 | 675 |

For GO-0.25, $E_2 \approx 10.5\ E_1$. For GO-3, $E_1$ is approximately 1.5 times lower than for GO-0.25, with $E_2 \approx 3.7\ E_1$ and $E_3 \approx 22\ E_1$. For GO-6 and GO-10, $E_1 \approx 2$–3 K is approximately an order of magnitude lower than for the low-GO compositions, with $E_2 \approx 19$–21 $E_1$ and $E_3 \approx (160$–$260)\ E_1$. Since $E_1 \approx 2$–3 K lies below the minimum measurement temperature of 4.5 K, its value should be interpreted as an effective fitting parameter with limited independent experimental constraint. A characteristic energy scale of approximately 27–50 K is found in all compositions, although it corresponds to different channel indices in the fitted representations. Higher characteristic energies of approximately 365–675 K are also obtained, represented by $E_2$ for GO-0.25 and $E_3$ for the other compositions. These results indicate that the measured thermal conductivity can be described by several effective activated contributions with distinct characteristic energy scales.

# 4. Discussion

## 4.1 Structure–property correlation

The structural data suggest substantial suppression of ordered GO stacking within the epoxy matrix, while no pronounced changes in the characteristic Raman vibrational bands of the epoxy network are detected up to GO-6. For compositions up to GO-6, XRD indicates an increase in the characteristic intermolecular spacing inferred from the main amorphous halo, whereas Raman spectroscopy shows an increasingly detectable contribution from the carbon phase. It should also be noted that the curing-agent-to-resin ratio varies among the investigated formulations. Consequently, changes in the epoxy-network structure associated with different curing-agent fractions may contribute to the observed thermal-transport behavior, and the effects of GO concentration cannot be completely isolated.

Against this background, the thermal-conductivity data reveal a qualitative change between GO-3 and GO-6: the low-temperature plateau in $\kappa(T)$ is no longer resolved for GO-6. In the adopted phenomenological description, this change is accompanied by a low-characteristic-energy Arrhenius-type contribution with $E_1 \approx 2$–3 K. These observations indicate a modification of the temperature dependence of heat transport rather than a simple change in its magnitude. Since $E_1$ lies below the minimum measurement temperature of 4.5 K, its precise value is subject to limited experimental constraints and should be regarded as an effective fitting parameter rather than direct evidence of a distinct low-energy excitation. The observed changes may be associated with modifications of the effective transport response of the heterogeneous matrix, including possible effects of structural disorder and GO–epoxy interfaces. However, the identification of specific interfacial or filler-related vibrational modes requires independent microscopic evidence.

At low GO content, the moderate increase in κ from GO-0.25 to GO-3 (12–21%) is accompanied by the appearance of a third, high-energy channel ($E_3$ = 600 K) in the phenomenological description, rather than by a uniform rescaling of $\kappa(T)$. The modest magnitude of this increase is consistent with the possibility of substantial interfacial thermal resistance at GO–epoxy boundaries [7, 8], although it does not independently establish the magnitude of this resistance or the absence of a percolating filler network. Moreover, $\kappa$ does not increase monotonically with GO content at high temperatures: near 300 K, GO-6 has a lower thermal conductivity than GO-3 (Fig. 5). The fitted characteristic energies span three approximate scales—a few kelvin, several tens of kelvin, and several hundred kelvin. These scales may reflect different effective contributions to thermal transport arising from the heterogeneous vibrational response of the polymer matrix, carbon domains, and their interfaces. However, a direct assignment to specific interfacial, molecular, or carbon-related vibrations cannot be made from the thermal-conductivity fits alone. In particular, the characteristic energies $E_i$ should not be directly identified with individual vibrational frequencies or interpreted as activation barriers for specific microscopic processes without independent spectroscopic or theoretical evidence.

### 4.2 Relation to the unified theory

In the heterogeneous epoxy–GO system, the activated channels should not be regarded as a direct microscopic identification of particular vibrational modes or of the coherence contribution in the Wigner formalism. Such an identification would require information on the vibrational spectrum, mode linewidths and mode-resolved transport coefficients. Nevertheless, the multichannel Arrhenius representation provides a consistent phenomenological description of the measured thermal conductivity. Within this description, increasing GO content modifies the relative amplitudes and characteristic energy scales of the effective transport contributions, while the functional form of the model remains unchanged.

### 4.3 Highest loading

GO-10 exhibits pronounced changes in its XRD response, including a shift of the main amorphous halo towards higher diffraction angles and an increase in its width. At the same time, its $\kappa(T)$ closely resembles that of GO-6. These observations suggest that the structural differences at the highest GO loading are not accompanied by a corresponding qualitative change in the measured thermal conductivity. Possible contributing factors include partial restacking or aggregation of the filler and differences in the epoxy-network composition associated with the substantially higher curing-agent fraction ($m_B/m_A$ = 54.33% for GO-10, compared with 20–30% for the lower-loading compositions). The available data do not allow these effects to be separated unambiguously, and the interpretation of the GO-10 structural anomalies therefore remains tentative.

## 5. Conclusions

The structural and thermal-transport properties of epoxy composites containing different concentrations of graphene oxide (GO) have been investigated to establish correlations between filler incorporation, structural organization, and heat conduction. XRD and Raman spectroscopy indicate substantial suppression of ordered GO stacking within the cured epoxy matrix, with no pronounced changes in the characteristic vibrational bands of the polymer network. An increase in the characteristic intermolecular spacing inferred from the main amorphous XRD halo is observed up to GO-6, whereas the structural changes at the highest loading (GO-10) are consistent with possible partial restacking or aggregation of the filler.

The thermal conductivity remains within the range typical of neat epoxy resins and exhibits predominantly glass-like behavior. Nevertheless, GO incorporation substantially modifies its temperature dependence. Over the investigated range of approximately 4.5–300 K, $\kappa(T)$ is described by a phenomenological model comprising a plateau contribution and two or three Arrhenius-type transport channels with characteristic energies spanning approximately 2–675 K. For GO-6 and GO-10, the low-temperature plateau is no longer resolved, and the fitted description includes a low-characteristic-energy contribution ($E_1 \approx 2–3$ K). These changes indicate a qualitative modification of the temperature dependence of heat transport rather than a simple conductivity enhancement.

Taken together, the results demonstrate that GO incorporation affects not only the magnitude of thermal conductivity but also the relative weights and characteristic energy scales of the effective contributions in its phenomenological representation. The modest, non-monotonic conductivity enhancement indicates that the thermal response cannot be understood solely in terms of an additional conducting filler phase. Although the microscopic origins of the activated contributions remain unresolved, the observed structure–property correlations provide an experimental basis for further investigations of vibrational dynamics, interfacial effects, and heat-transport mechanisms in disordered polymer–carbon composites.

**Acknowledgments**

The work of A.I.K., V.V.S, Yu.V.H. and S.V.C. was supported by the National Research Foundation of Ukraine (Project No 2023.03/0012). The work of A.J. and D.S. was supported by National Science Centre, Poland (Grant No. 2022/45/B/ST3/02326).

**Conflict of interest**

The authors declare no conflict of interest.

**Data accessibility statement**

The data that support the findings of this study are available upon reasonable request from the authors.

**References**

1. Chen H., Ginzburg V. V., Yang J., Yang Y., Liu W., Huang Y., Du L., Chen B. Thermal conductivity of polymer-based composites: Fundamentals and applications. *Progress in Polymer Science*. 2016. Vol. 59. P. 41–85. DOI: https://doi.org/10.1016/j.progpolymsci.2016.03.001.
2. Li A., Zhang C., Zhang Y.-F. Thermal conductivity of graphene-polymer composites: Mechanisms, properties, and applications. *Polymers*. 2017. Vol. 9, No. 9. Article 437. DOI: https://doi.org/10.3390/polym9090437.

3. Balandin A. A. Thermal properties of graphene and nanostructured carbon materials. *Nature Materials*. 2011. Vol. 10. P. 569–581. DOI: https://doi.org/10.1038/nmat3064.
4. Renteria J. D., Nika D. L., Balandin A. A. Graphene thermal properties: Applications in thermal management and energy storage. *Applied Sciences*. 2014. Vol. 4, No. 4. P. 525–547. DOI: https://doi.org/10.3390/app4040525.
5. Im H., Kim J. Thermal conductivity of a graphene oxide–carbon nanotube hybrid/epoxy composite. *Carbon*. 2012. Vol. 50. P. 5429–5440. DOI: https://doi.org/10.1016/j.carbon.2012.07.029.
6. Kim H., Abdala A. A., Macosko C. W. Graphene/polymer nanocomposites. *Macromolecules*. 2010. Vol. 43. P. 6515–6530. DOI: https://doi.org/10.1021/ma100572e.
7. Danayat S., Nayal A. S., Tarannum F., Annam R., Muthaiah R., Arulanandam M. K., Garg J. Superior enhancement in thermal conductivity of epoxy/graphene nanocomposites through use of dimethylformamide (DMF) relative to acetone as solvent. *MethodsX*. 2023. Vol. 11. Article 102319. DOI: https://doi.org/10.1016/j.mex.2023.102319.
8. Wang Y., Yang C., Pei Q.-X., Zhang Y. Some aspects of thermal transport across the interface between graphene and epoxy in nanocomposites. *ACS Applied Materials & Interfaces*. 2016. Vol. 8. P. 8272–8279. DOI: https://doi.org/10.1021/acsami.6b00325.
9. Ziman J. M. *Electrons and Phonons: The Theory of Transport Phenomena in Solids*. Oxford : Clarendon Press, 1960.
10. Cahill D. G., Pohl R. O. Thermal conductivity of amorphous solids above the plateau. *Physical Review B*. 1987. Vol. 35. P. 4067–4073. DOI: https://doi.org/10.1103/PhysRevB.35.4067.
11. Isaeva L., Barbalinardo G., Donadio D., Baroni S. Modeling heat transport in crystals and glasses from a unified lattice-dynamical approach. *Nature Communications*. 2019. Vol. 10. Article 3853. DOI: https://doi.org/10.1038/s41467-019-11572-4.
12. Allen P. B., Feldman J. L. Thermal conductivity of glasses: Theory and application to amorphous Si. *Physical Review Letters*. 1989. Vol. 62. P. 645–648. DOI: https://doi.org/10.1103/PhysRevLett.62.645.
13. Allen P. B., Feldman J. L., Fabian J., Wooten F. Diffusons, locons and propagons: Character of atomic vibrations in amorphous Si. *Philosophical Magazine B*. 1999. Vol. 79. P. 1715–1731. DOI: https://doi.org/10.1080/13642819908223054.
14. Simoncelli M., Marzari N., Mauri F. Unified theory of thermal transport in crystals and glasses. *Nature Physics*. 2019. Vol. 15. P. 809–813. DOI: https://doi.org/10.1038/s41567-019-0520-x.
15. Simoncelli M., Marzari N., Mauri F. Wigner formulation of thermal transport in solids. *Physical Review X*. 2022. Vol. 12. Article 041011. DOI: https://doi.org/10.1103/PhysRevX.12.041011.

16. Simoncelli M., Mauri F., Marzari N. Thermal conductivity of glasses: First-principles theory and applications. *npj Computational Materials*. 2023. Vol. 9. Article 106. DOI: https://doi.org/10.1038/s41524-023-01033-4.
17. Di Lucente E., Simoncelli M., Marzari N. Crossover from Boltzmann to Wigner thermal transport in thermoelectric skutterudites. *Physical Review Research*. 2023. Vol. 5. Article 033125. DOI: https://doi.org/10.1103/PhysRevResearch.5.033125.
18. Legenstein L., Reicht L., Wieser S., Simoncelli M., Zojer E. Heat transport in crystalline organic semiconductors: Coexistence of phonon propagation and tunneling. *npj Computational Materials*. 2025. Vol. 11. Article 29. DOI: https://doi.org/10.1038/s41524-025-01514-8.
19. Krivchikov A., Andersson O., Korolyuk O., Kryvchikov O. Thermal conductivity of solid triphenyl phosphite. *Molecules*. 2022. Vol. 27, No. 23. Article 8399. DOI: https://doi.org/10.3390/molecules27238399.
20. Krivchikov A. I., Horbatenko Y. V., Korolyuk O. A., Romantsova O. O., Kryvchikov O. O., Szewczyk D., Jeżowski A. Exponential approximation of the coherence contribution to the thermal conductivity of complex clathrate-type crystals. *Materialia*. 2023. Vol. 32. Article 101944. DOI: https://doi.org/10.1016/j.mtla.2023.101944.
21. Krivchikov A. I., Korolyuk O. A. Empirical universal approach to describing the thermal conductivity of amorphous polymers: Effects of pressure, radiation and the Meyer–Neldel rule. *Low Temperature Physics*. 2024. Vol. 50. P. 328–341. DOI: https://doi.org/10.1063/10.0025299.
22. Krivchikov A. I., Jeżowski A., Konstantinov V. A., Sagan V. V., Korolyuk O. A., Szewczyk D. Enhancing thermal transport in ABS polymer with graphene oxide: Insights into low-temperature thermal conductivity behavior and correlation with boson peak anomaly. *Thermochimica Acta*. 2024. Vol. 733. Article 179696. DOI: https://doi.org/10.1016/j.tca.2024.179696.
23. Nataj Z. E., Xu Y., Wright D., Brown J. O., Garg J., Chen X., Kargar F., Balandin A. A. Cryogenic characteristics of graphene composites—evolution from thermal conductors to thermal insulators. *Nature Communications*. 2023. Vol. 14. Article 3190. DOI: https://doi.org/10.1038/s41467-023-38508-3.
24. Dolbin A. V., Khlistyuck M. V., Esel'son V. B., Gavrilko V. G., Vinnikov N. A., Basnukaeva R. M., Maluenda I., Maser W. K., Benito A. M. The effect of the thermal reduction temperature on the structure and sorption capacity of reduced graphene oxide materials. *Applied Surface Science*. 2016. Vol. 361. P. 213–220. DOI: https://doi.org/10.1016/j.apsusc.2015.11.167.
25. Dolbin A. V., Vinnikov N. A., Esel'son V. B., Gavrilko V. G., Basnukaeva R. M., Khlistyuck M. V., Maser W. K., Benito A. M. The effect of graphene oxide reduction temperature on the kinetics of low-temperature sorption of hydrogen. *Low Temperature Physics*. 2019. Vol. 45. P. 422–426. DOI: https://doi.org/10.1063/1.5093523.

26. Jeżowski A., Mucha J., Pompe G. Thermal conductivity of the amorphous alloy $Fe_{40}Ni_{40}P_{14}B_6$ between 80 and 300 K. *Journal of Physics D: Applied Physics*. 1987. Vol. 20. P. 1500–1506. DOI: https://doi.org/10.1088/0022-3727/20/11/022.
27. Harper A. F., Iwanowski K., Witt W. C., Payne M. C., Simoncelli M. Vibrational and thermal properties of amorphous alumina from first principles. *Physical Review Materials*. 2024. Vol. 8, No. 4. Article 043601. DOI: https://doi.org/10.1103/PhysRevMaterials.8.043601.
28. Simoncelli M., Fournier D., Marangolo M., Balan E., Béneut K., Baptiste B., Doisneau B., Marzari N., Mauri F. Temperature-invariant crystal–glass heat conduction: From meteorites to refractories. *Proceedings of the National Academy of Sciences*. 2025. Vol. 122, No. 28. Article e2422763122. DOI: https://doi.org/10.1073/pnas.2422763122.
29. Anderson P. W., Halperin B. I., Varma C. M. Anomalous low-temperature thermal properties of glasses and spin glasses. *Philosophical Magazine*. 1972. Vol. 25, No. 1. P. 1–9. DOI: https://doi.org/10.1080/14786437208229210.
30. Phillips W. A. Tunneling states in amorphous solids. *Journal of Low Temperature Physics*. 1972. Vol. 7, No. 3–4. P. 351–360. DOI: https://doi.org/10.1007/BF00660072.
31. Zeller R. C., Pohl R. O. Thermal conductivity and specific heat of noncrystalline solids. *Physical Review B*. 1971. Vol. 4. P. 2029–2041. DOI: https://doi.org/10.1103/PhysRevB.4.2029.
32. *Low-Temperature Thermal and Vibrational Properties of Disordered Solids: A Half-Century of Universal “Anomalies” of Glasses* / ed. by M. A. Ramos. World Scientific Publishing Europe, 2022. 504 p. ISBN 978-1-80061-257-0.
33. Peierls R. Zur kinetischen Theorie der Wärmeleitung in Kristallen. *Annalen der Physik*. 1929. Vol. 395. P. 1055–1101. DOI: https://doi.org/10.1002/andp.19293950803.
34. Karpov V. G., Klinger M. I., Ignatiev P. N. Atomic tunneling states and low-temperature anomalies of thermal properties in amorphous materials. *Solid State Communications*. 1982. Vol. 44, No. 3. P. 333–337. DOI: https://doi.org/10.1016/0038-1098(82)90866-3.
35. Buchenau U., Galperin Yu. M., Gurevich V. L., Parshin D. A., Ramos M. A., Schober H. R. Interaction of soft modes and sound waves in glasses. *Physical Review B*. 1992. Vol. 46, No. 5. P. 2798–2808. DOI: https://doi.org/10.1103/PhysRevB.46.2798.
36. Ramos M. A., Buchenau U. Low-temperature thermal conductivity of glasses within the soft-potential model. *Physical Review B*. 1997. Vol. 55, No. 9. P. 5749–5754. DOI: https://doi.org/10.1103/PhysRevB.55.5749.
37. Horbatenko Yu. V., Sagan V. V., Korolyuk O. A., Romantsova O. O., Krivchikov A. I. Temperature dependences of thermal conductivity of solid heterogeneous crystalline and amorphous materials: An empirical approach to the description in the high-temperature region. *Low Temperature Physics*. 2024. Vol. 50, No. 5. P. 379–388. DOI: https://doi.org/10.1063/10.0025621.

38. Korolyuk O. A., Sagan V. V., Krivchikov A. I., Konstantinov V. A., Horbatenko Yu. V. Multi-channel heat transfer in molecular crystals and their solutions (Review article). *Low Temperature Physics*. 2025. Vol. 51, No. 9. P. 1048–1060. DOI: https://doi.org/10.1063/10.0038988.
39. Korolyuk O. A., Krivchikov A. I., Horbatenko Yu. V., Tanguy A., Giordano V. M., Szewczyk D., Jeżowski A., Nakazawa Y. Bridging glass-like and crystal-like thermal transport: A unified model for disordered and complex solids. *Journal of Non-Crystalline Solids*. 2026. Vol. 690. Article 124273. DOI: https://doi.org/10.1016/j.jnoncrysol.2026.124273.
40. Berman R. *Thermal Conduction in Solids*. Oxford : Clarendon Press, 1976.
41. Jeżowski A., Krivchikov A. I., Korolyuk O. A., Sagan V. V., Horbatenko Y. V., Cherednychenko S. V., Dolbin A. V., Vinnikov N. A., Bezkrovnyi O., Szewczyk D. Universal multichannel coherence-mediated thermal transport in carbon allotropes: Experimental evidence and scaling laws. *arXiv*. 2026. arXiv:2611.14165. URL: https://arxiv.org/abs/2611.14165.
42. Garrett K. W., Rosenberg H. M. The thermal conductivity of epoxy-resin / powder composite materials. *Journal of Physics D: Applied Physics*. 1974. Vol. 7, No. 9. P. 1247–1258. DOI: https://doi.org/10.1088/0022-3727/7/9/311.
43. Scheibner W., Jäckel M. Thermal conductivity and specific heat of an epoxy resin/epoxy resin composite material at low temperatures. *physica status solidi (a)*. 1985. Vol. 87, No. 2. P. 543–547. DOI: https://doi.org/10.1002/pssa.2210870216.
44. Nicholls C. I., Rosenberg H. M. The excitation spectrum of epoxy resins; neutron diffraction, specific heat and thermal conductivity at low temperatures. *Journal of Physics C: Solid State Physics*. 1984. Vol. 17, No. 7. P. 1165–1178. DOI: https://doi.org/10.1088/0022-3719/17/7/011.
45. Xie Y., Yuan P., Wang T., Hashemi N., Wang X. Switch on the high thermal conductivity of graphene paper. *Nanoscale*. 2016. Vol. 8, No. 40. P. 17581–17597. DOI: https://doi.org/10.1039/C6NR06402G.
46. Chen T., Huang Y., Wei L., Xu T., Xie Y. Thermal and electrical transport in partly-reduced graphene oxide films: The effect of low temperature and structure domain size. *Carbon*. 2023. Vol. 203. P. 130–140. DOI: https://doi.org/10.1016/j.carbon.2022.11.051.